\documentclass[5p,authoryear,10pt]{elsarticle}
\usepackage{mathrsfs}
\usepackage{graphicx}
\usepackage{amsmath}
\usepackage{amssymb}
\usepackage{amsthm}
\usepackage{amsfonts}
\usepackage{natbib}
\usepackage{hyperref}
\hypersetup{colorlinks=true,
citecolor=green,
linkcolor=red,
urlcolor=cyan,
pdftex}
\journal {BioSystems. Licence \href{https://creativecommons.org/licenses/by-nc-nd/4.0/}{CC-BY-NC-ND~4.0}.}
\newcommand{\can}{\citet}
\newcommand{\dsty}{\displaystyle}

\newcommand{\sss}{\scriptscriptstyle}

\newcommand{\eq}{equation}

\newcommand{\lra}{\leftrightarrow}

\newcommand{\be}{\begin{\eq}}
\newcommand{\ee}{\end{\eq}}

\newcommand{\arr}{array}
\newcommand{\ba}{\begin{\arr}}
\newcommand{\ea}{\end{\arr}}

\newtheorem{law}{Law}
\newtheorem{defin}{Definition}

\begin{document}
\title{Memory retrieval dynamics and storage capacity of a modular \\ network model of association cortex with featural decomposition}
\author{Carlo~Fulvi~Mari\hspace{.1cm} \corref{*}}
\date{Submitted 30/4/2021 - Revised 2/9/2021 - Accepted 31/11/2021}

\cortext[*]{E-mail address: \href{mailto:cfmphys@gmail.com}{cfmphys@gmail.com}
\\ {\indent \href{https://orcid.org/0000-0002-5828-9412}{\includegraphics[width=3mm]{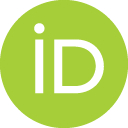} https://orcid.org/0000-0002-5828-9412}}}

\begin{abstract}
The primate heteromodal cortex presents an evident functional modularity at a mesoscopic level, with physiological and anatomical evidence pointing to it as likely substrate of long-term memory. In order to investigate some of its properties, a model of multimodular autoassociator is studied. Each of the many modules represents a neocortical functional ensemble of recurrently connected neurons and operates as a Hebbian autoassociator, storing a number of local {\it features} which it can recall upon cue. The global memory {\it patterns} are made of combinations of features sparsely distributed across the modules. Intermodular connections are modelled as a finite-connectivity random graph. Any pair of features in any respective pair of modules is allowed to be involved in several memory patterns; the coarse-grained modular network dynamics is defined in such a way as to overcome the consequent ambiguity of associations. Effects of long-range homeostatic synaptic scaling on network performance are also assessed.

The dynamical process of cued retrieval almost saturates a natural upper bound while producing negligible spurious activation. The extent of finite-size effects on storage capacity is quantitatively evaluated. In the limit of infinite size, the functional relationship between storage capacity and number of features per module reduces to that which other authors found by methods from equilibrium statistical mechanics, which suggests that the origin of the functional form is of a combinatorial nature. In contrast with its apparent inevitability at intramodular level, long-range synaptic scaling results to be of minor relevance to both retrieval and storage capacity, casting doubt on its existence in the neocortex. A conjecture is also posited about how statistical fluctuation of connectivity across the network may underpin spontaneous emergence of semantic hierarchies through learning.
\end{abstract} 


\maketitle
{\noindent \footnotesize {\it Ref.}: C. Fulvi Mari, {\it BioSystems}, 104570 (2022) \\ {\it DOI:} \href{https://doi.org/10.1016/j.biosystems.2021.104570}{10.1016/j.biosystems.2021.104570}}\\
\noindent\rule{6.5cm}{.5pt}

\section{Introduction}
Although no definition of archetypical elementary module of the primate neocortex is universally agreed, there is little doubt that a functional processing unit does exist, even if its anatomical boundaries may be not always well defined or very evident. Such functional unit is generally understood to extend vertically through the six cortical layers, to have a base area of the order of 1~\!mm$^{2}$, and to contain a number of neurons in the order of 10$^{5}$; while in some areas it may resemble a column, in other areas it has a more elongated and less regular horizontal section. The size of columns as defined by afferent long-range axonal bundles seems to only depend weakly on the size of the cortex across species, suggesting that phylogenetic increase in cortical surface area correlates with increase in number of functional ensembles rather than with their individual size \citep{Bugbee1983}. In the following, a generic unit, called {\sl module}, is defined in terms of its functional properties, refraining from any immediate cytoarchitectonic correspondence. The neurons inside each module are henceforth assumed to be densely interconnected and to also project axons to neurons of other modules. 

The recurrent connectivity as well as the experimental evidence of reverberating activity have led to the formulation of mathematical models with the aim of understanding, inter alia, how the very numerous modules are organised in terms of architecture of connections and information flow, processing, storage and retrieval. The ultimate objective is, of course, to uncover the mechanistic neuronal underpinning of cognitive processes, which invariably involve large scale structures and therefore a multitude of modules. The attention of the present work is especially on the storage and retrieval processes thought to take place in the heteromodal areas of the cortex that are chiefly responsible for semantic memory \citep{Binder2009}. A brief overview of the anatomical and physiological background of the model, as well as its position among the theoretical models already developed, can be found in the Discussion. The rest of this Section is dedicated to outlining the present work.

Each module is modelled as an autoassociator that can store a large number of local memories, called {\sl features}. Neurons of any module are assumed to project axons to neurons of other modules, the axonal projections from each module being concentrated onto a relatively small number of other modules, as if following the edges of an underlying dilute random graph. The intermodular, or {\sl long-range}, synapses as well as the intramodular ones are supposed to be Hebbian, so that the large number of modules can function as a multimodular autoassociator. Each global memory {\sl pattern} of the network is hence made of a combination of local features, specific to the pattern, which are represented in a sparse number of modules, the other modules being in a quiescent state. Different patterns involve different combinations of features encoded in different sets of modules, although patterns can share features. The respective states of activation of interconnected modules are positively correlated across the set of memory patterns, which may be related to semantic value. During a putative learning phase, new features may be learned in any module and new Hebbian associations between new or already learned features in different modules may be formed. As any pair of features in respective adjacent modules can be simultaneously involved in more than one pattern, the reactivation of any feature in any module may not determine univocally the features that neighbouring modules should reproduce ({\sl local ambiguity}). 

Neurons are not represented individually; rather, in order to allow for modelling the dynamics of a network with a large number of modules, a coarse-graining approach is adopted which is founded on previous results by the author and others (cf. Sections \ref{model} and \ref{theories}).

The featural decomposition of memory patterns is biologically and cognitively meaningful. In particular, the sharing of features across patterns may hold semantic value and it can also be argued that it may facilitate important cognitive abilities (e.g., generalisation, analogies and similarities, recognition invariance, statistical inference). A further advantage of functional modularity may concern the economy of memory resources; indeed, if features are sufficiently elementary and, at the same time, not too specific, so to occur in many patterns, they do not need to be learned more than once. All these properties would be hard to achieve in a unimodular, (statistically) homogeneous autoassociator, whereby any new pattern is stored independently.

An accurate balance, possibly by neuromodulation, between the intramodular recurrent processes and the intermodular interactions is needed in order to take advantage of both local featural processes and global cooperation while avoiding to fall into dysfunctional states in which long-range interactions are so strong as to suppress local ones or, conversely, each module is so weakly dependent on the activity of the others that the system may get trapped in a meaningless patchwork of pieces from several memories, known as {\sl memory glass} \citep{OKT,FMT,CFM2000,CFM2004}. The dynamics is appropriately defined to overcome these hindrances while being biologically plausible. The network performance is evaluated in terms of cued memory retrieval and storage capacity.

Because features are activated potentially by many patterns during learning, it seems most likely that some homeostatic scaling mechanism intervenes to keep every module in a reasonable working range. In fact, the need for a mechanism of synaptic weight scaling in (unimodular) Hebbian networks has been evident since early theoretical works. Forms of synaptic homeostasis have already been experimentally found and it is not too difficult to envisage what kind of intramodular processes may underpin that function. It seems instead more complex to devise a mechanism that produce functional scaling of Hebbian synapses between different cortical modules, for neurons of different modules are scattered across a relatively wide range of the cortex and recruited simultaneously less frequently than neurons within the same module. As any pair of features in adjacent modules is shared by different patterns much less often during learning than any feature in any module, it is possible that long-range scaling may be dispensable; however, given the potential of inter-modular associative synapses to cause percolating spread of incorrect activity, this qualitative argument cannot be conclusive. In this work an attempt is made to evaluate quantitatively the effect of the presence or absence of long-range synaptic scaling on memory storage capacity and cued retrieval.

The choice was made to study out-of-equilibrium dynamics. In general, this allows for delving into the details of activity spread and for not neglecting events with too small a probability rate to happen in a cognitively relevant timeframe, which would instead elude the more common approaches of equilibrium statistical mechanics. As a simple example one may consider the persistence of activity in a wrongly activated module. Commonly adopted stochastic dynamics (irreducible reversible Markov chains) would allow for a positive probability rate of spontaneous decay to quiescence. An equilibrium study, which makes use of the stationary distribution of the stochastic dynamics, would have implicitly let such an event happen an arbitrarily large number of times; however, the typical timescale for the decay could be much larger than the timescale of a cognitive window, during which that event may be unlikely to happen. Of course, the present approach has its own drawbacks, first among which the lesser amenability to mathematical analysis.

The main advancements of the present research with respect to previous results \citep{CFM2004} are: (1) explicit assignment of numerous local features to each module and actual storage of the relative multiple intermodular featural associations (instead of only storing the features of the pattern to retrieve and relative associations while modelling the presence of the others statistically); (2) heuristic and simulation-based estimate of the memory storage capacity of the multimodular network, including quantification of finite-size effects (instead of only the single-module-based signal-to-noise capacity analysis and retrieval simulations for very large networks); (3) evaluation of the potential relevance of homeostatic scaling of long-range synapses (not present, nor feasible, in previous work).

\section{Model network and dynamics} \label{model}
The dynamical unit is the module of neurons: it is assumed to function as a Hebbian autoassociator that can store and retrieve $F$ neuronal (local) activity configurations, called {\it features}. The network is composed of a large number $M$ of modules, that interact through a net of (quenched) random links, and stores $P$ global patterns. The dynamics is defined at a coarse-grained modular level, which has foundation on results obtained at the neuronal network level by statistical-mechanical \citep{Parga} and signal-to-noise \citep{CFM2004} analyses of biologi\-cally-realistic multimodular models. If one accepts the hypothesis that functional modules exist which store cognitive features, as quite convincingly supported by experimental evidence, coarse-graining allows for meaningful quantitative modelling and large scale simulations of the processes that take place at the mesoscopic level and that underpin cognitive functions (in this case, memory functions). A table of symbols and acronyms that recur in the following is in Appendix~B.

\begin{figure}[t]
\includegraphics[height=6cm]{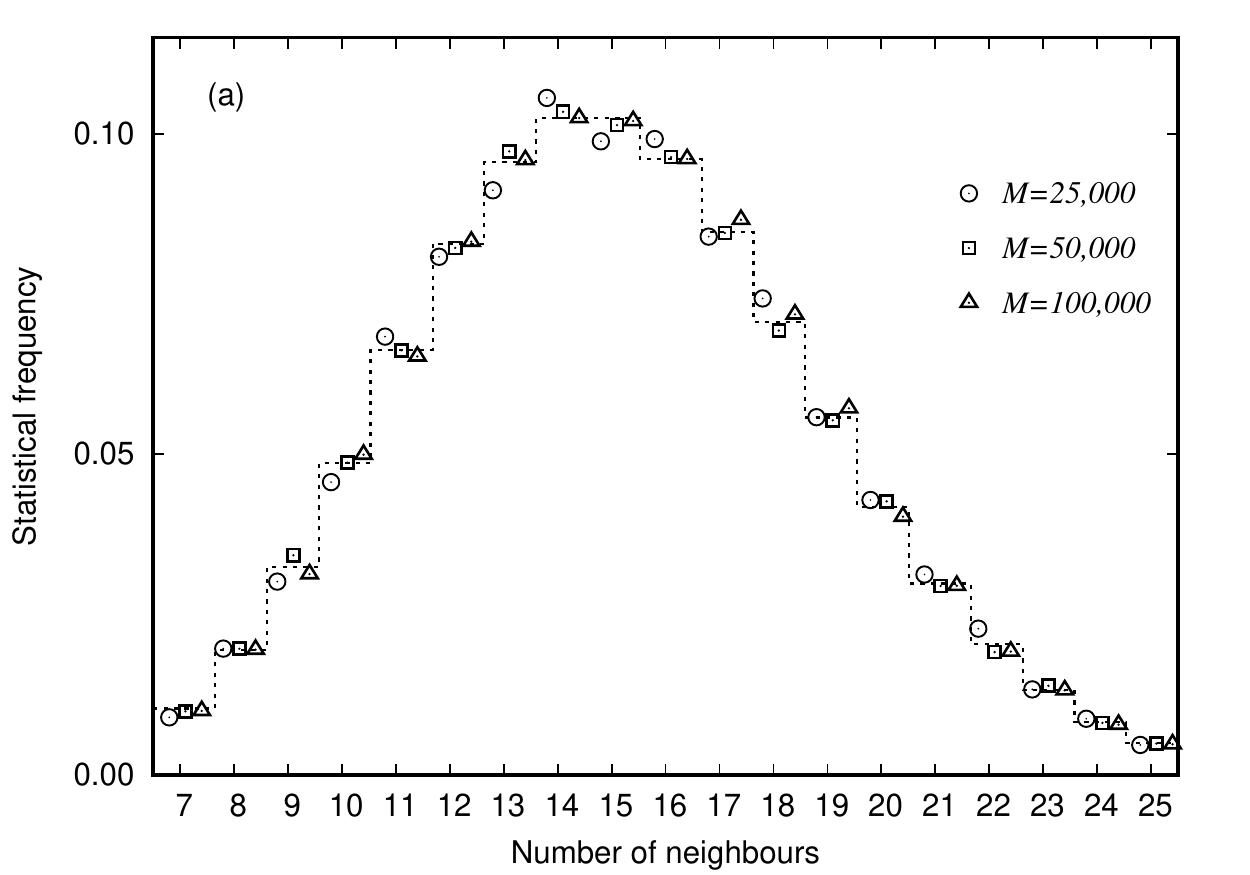}
\caption{\small{Statistics of model connections: Frequency of nodes as a function of number of neighbours across the graph, the data coming from 3 random graphs of different sizes and the histogram reporting the expected binomial distribution.}
\label{gracstat}}
\end{figure}

\subsection{Architecture}
The connections between modules are constructed as in a large, extremely dilute undirected simple (i.e., without 1-loops or multiple edges) binomial random graph \citep{Gilbert1959,ErdosRenyi1960,BollobasII}: the nodes of the graph represent the modules while the edges represent the connections between modules. The probability for any pair of nodes to be joined by an edge is $z/M$, with $z$ constant, so that the average number of neighbours across modules remains about constant at large, increasing $M$. As in standard terminology, modules that are connected to each other are called adjacent and the number of neighbours of any given module is called coordination number of the latter, $z$ then being called the mean coordination number. It should be noticed that adjacency in graph-distance does not reflect geometric closeness and, therefore, any two adjacent nodes may be representing cortical modules that are topographically distant. A highly diluted graph model (edge probability $\propto \ln\!M /M)$ could also be used that would guarantee the graph to be connected almost surely in the infinite-size limit; however, with biologically-realistic finite sizes, both graph models are reasonable approximations.

Experimental evidence shows that reciprocal interco\-lumnar connections in heteromodal cortex are far more frequent than they would be if the connections in the two directions were built independently of each other \citep{GoldmanRakic,Pucak1996,Ichinohe2012}. It was therefore decided that a model with reciprocal intermodular connection, albeit still being approximate, is better for multimodular networks than a model with independent directed edges. The best approximation would most likely sit somewhere in between the two extremes, but it would also be less amenable to mathematical and, to a lesser extent, simulation study. Besides, and perhaps more importantly, it seems reasonable that semantically related modules be more likely to interact reciprocally, to the advantage of memory processes.

Figure~\ref{gracstat} shows the frequency of nodes in the network as a function of the number of neighbours for three realisations of the graph, with respectively 25,000, 50,000, and 100,000 modules, and the corresponding binomial distribution in dashed histogram. Assuming that a functional column of the neocortical association areas has a horizontal section of about 1~\!mm$^2$, the simulations described below would then represent 250-1,000~\!cm$^2$ of the cortical sheet, equivalent to about 2.5-10\% of the total, which seems realistic \citep{Binder2009}. Deviations of the mean number of neighbours from the expected value predictably decrease in amplitude with larger networks; although being quite small already for the smallest size, they are not negligible in any of them and, together with the randomness of the featural decomposition, they cause visible variability in network behaviour. 

\subsection{Memory patterns and features}
\begin{figure}[t]
\includegraphics[height=6cm]{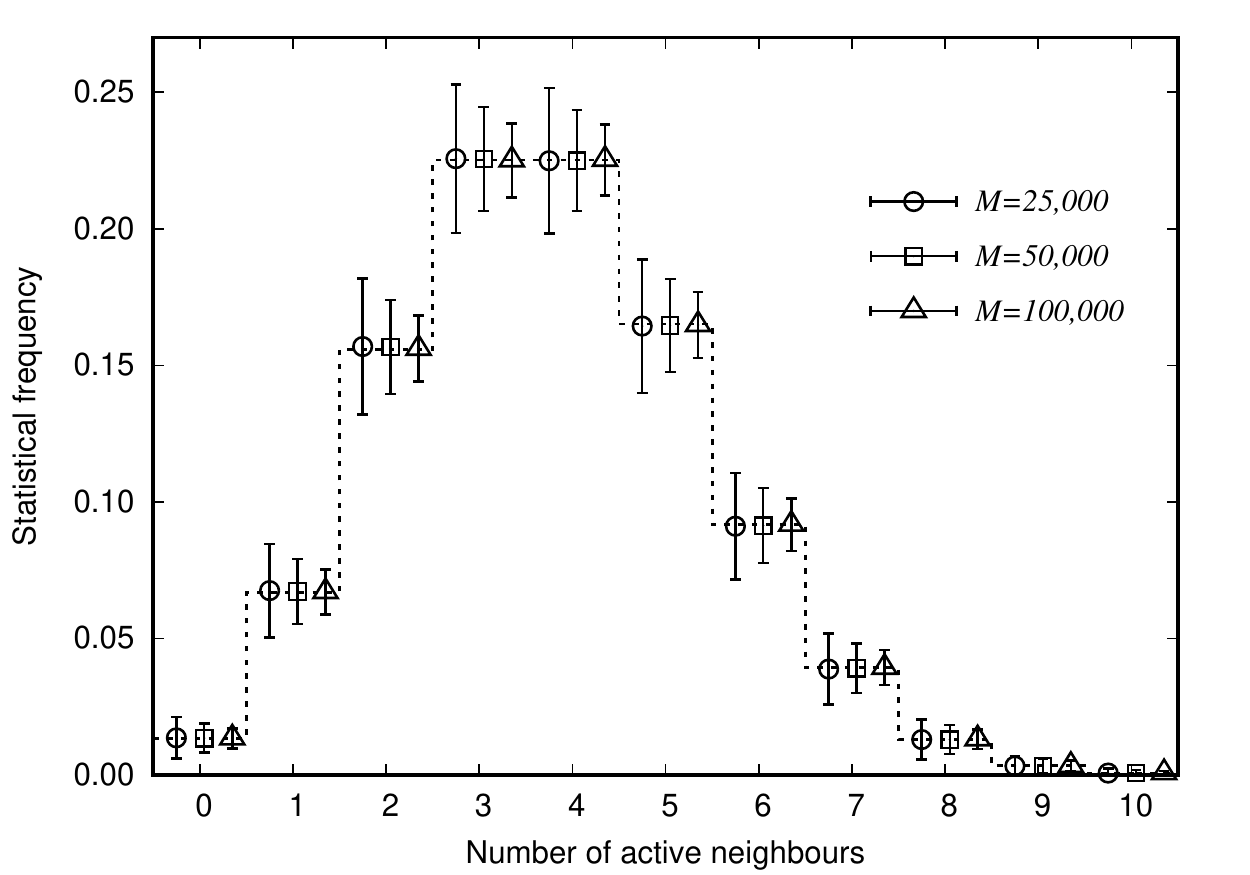}
\caption{\small{Statistics of model activity patterns: Frequency of active neighbours of active modules that have each in total 15 neighbours, in samples of 20,000 patterns for each of 3 network sizes (mean and variance stabilise after a few thousands data-points; each error bar is twice as long as the sample standard deviation).}
\label{actstat}}
\end{figure}

Memory patterns are made of specific combinations of features stored in respective subsets of modules. For each pattern to be stored, the modules that are involved by that pattern are chosen randomly but not independently: on average, any two modules are more likely be recruited simultaneously if they are adjacent, while they are otherwise recruited independently, in such a way as to keep the network mean activity about the same for all patterns. Summarily, for any modules ${\mathbf A}$ and ${\mathbf B}$, the average marginal probabilities are given by
\be
\left\{
\ba{l}
{\mathbb{P}}\left({\mathbf A} \,active\right)=\tau\\
\\
{\mathbb{P}}\left({\mathbf A} \,active, {\mathbf B}\,active\mid {\mathbf A}\lra {\mathbf B}\right)=\tau t_{1}\\
\\
{\mathbb{P}}\left({\mathbf A} \,quiescent, {\mathbf B}\,active\mid {\mathbf A}\lra {\mathbf B}\right)=\left(1-\tau\right) t_{0}\\
\\
{\mathbb{P}}\left({\mathbf A} \,active, {\mathbf B}\,active\mid {\mathbf A}\,\not\!\lra {\mathbf B}\right)=\\[.25cm]
\hspace*{1.5cm} ={\mathbb{P}}\left({\mathbf A} \,active\right)\,{\mathbb{P}}\left({\mathbf B}\,active\right)=\tau^{2},
\ea
\right.
\label{scheme}
\ee
one of the consistency requirements being
\be
(1-\tau)\,t_{0}=(1-t_{1})\,\tau.
\label{consist}
\ee
This probabilistic scheme was studied more in detail, mathematically and numerically, in \can{CFM2000}, where its consistency (and, therefore, physical meaningfulness) was proven to hold in a biologically plausible range of the parameters, and was used in \can{FMT} to suppress memory-glass states. The actual realisation of memory patterns with the correlated statistics is here achieved by numerical implementation of a purposely defined stochastic (heat-bath) process \citep{CFM2000}. In the following, the active modules in any memory pattern will be said to constitute the {\it foreground} (FG) of that pattern, while the quiescent modules will be said to constitute the {\it background} (BG). The statistics of active neighbours clearly depends on the total number of neighbours; as an example, Fig.~\ref{actstat} shows the distribution of the number of active neighbours of FG modules that have a total of 15 neighbours each, together with the expected histogram.

For each module, to each of the $P$ global patterns that recruit that module a local feature is then assigned, which is randomly chosen (with uniform p.d.f.) out of the $F$ features stored in the module. This completes the construction of the set of memory patterns.

\subsection{Memory storage and homeostatic synaptic scaling}
Associations between features in adjacent modules are assumed to have been stored during the learning stage according to a Hebbian-like rule. If ${\mathbf A}$ and ${\mathbf B}$ are adjacent modules and the presented pattern recruits feature $a$ in ${\mathbf A}$ and feature $b$ in ${\mathbf B}$, with $a, b \in [F]$, an association is created specifically between the two features, formally represented as a unitary increment from null baseline. If another pattern is then presented that recruits the same two features, two possibilities are considered: (1) the weight is incremented by a further unit, or (2) there is no further increment, as the association already exists. In both cases there is never decrement of the weight, which differs in principle from a covariance rule and does not permit feature-specific inhibition. Case (1) corresponds to the hypothesis according to which synaptic scaling of long-range contacts is not present (nLRSS network), while in case (2) it is implemented (LRSS network).

It is assumed that repeated activation of a feature during learning does not engender further Hebbian increments of the synaptic weights between neurons of the same module, as a consequence of short-range synaptic scaling.\,There\-fore, it is as if any memory feature would only be learned once, while the weights of associations between features in different modules can be repeatedly incremented or not, depending on absence or presence of LRSS. Both cases will be taken into account and the respective results then compared.

\subsection{Network dynamics} \label{dynamics}
If a portion of a stored memory pattern is presented to the quiescent network as a cue, it is a requirement that the dynamics drive the network to retrieve the rest of that memory pattern. Besides, there should be no significant activation of features that do not belong in that pattern. By significant it is meant not only that the fraction of network that is activated be not significantly larger than what would correspond to the exact reactivation of the pattern, but also that any set of wrongly activated features have not significant similarity (i.e., overlap) with any of the other stored patterns. The cue will always be considered as made of complete elementary features rather than individual neuronal firing-rates, in the assumption that any module is capable of quick completion of the local feature when the latter is not fully represented in the cue.

All that follows below can be quite easily formalised mathematically; however, this would unavoidably require the introduction of many symbols, often with multiple indices, which would be cumbersome without being necessary for the understanding or reproducibility of model and results. Therefore, it was preferred to adopt a style of presentation as informal as possible.

In order to make the description of the dynamical properties of the network more concise, it is useful to first establish some definitions:
\begin{defin}[\it Association weight]
\rm Let ${\mathbf A}$ and ${\mathbf B}$ be any two active adjacent modules, which are reproducing respectively features $a$ and $b$. In the nLRSS network, the {\bf association weight of} the pair $(a,b)$ is defined as the number of memory patterns that recruit simultaneously those two features. In the LRSS network, the association weight of the pair $(a,b)$ is equal to 1 if at least one of the memory patterns recruits simultaneously those two features, while it is equal to 0 otherwise. In both cases, the {\bf association weight onto} any feature $a$ in ${\mathbf A}$ is equal to the association weight of the pair $(a,b)$. If the association weight of $(a,b)$ is positive, module ${\mathbf A}$ is said to {\bf support} module ${\mathbf B}$ (and vice versa).
\label{aw}
\end{defin}
\begin{defin}[\it Compound association weight]
\rm Given any module ${\mathbf A}$ and its $n$ active neighbours $({\mathbf B}_{k})_{\sss{k\in[n]}}$, which are respectively reproducing the features $(b_{k})_{\sss{k\in[n]}}$, the compound association weight onto any feature $a$ in ${\mathbf A}$ is the sum of the association weights of all the pairs $(a,b_{k})_{\sss{k\in[n]}}$.
\label{caw}
\end{defin}

As an illustrative example, consider the three modules ${\mathbf A}$, ${\mathbf B}_{1}$, ${\mathbf B}_{2}$, with ${\mathbf B}_{1}$ and ${\mathbf B}_{2}$ being neighbours of ${\mathbf A}$ (that is, $\mathbf{B_{1}\lra A\lra B_{2}}$) and reproducing, respectively, features $b_{1}$ and $b_{2}$. Assume that $n_{1}>0$ patterns recruit the pair $(a,b_{1})$ and that $n_{2}>0$ patterns recruit the pair $(a,b_{2})$. In the nLRSS network, the association weight of the pair $(a,b_{1})$ is $n_{1}$, while that for the pair $(a,b_{2})$ is $n_{2}$; the compound association weight onto feature $a$ is then equal to $n_{1}+n_{2}$. In the LRSS network, the association weights of the two pairs $(a,b_{1})$ and $(a,b_{2})$ are both equal to 1, and the compound association weight onto feature $a$ is then equal to 2.

The laws of the dynamics are defined in such a way as to exploit combinatorial properties of memory storage in order to make incorrect activation unlikely enough. They are:
\begin{law}[\it Oscillations]
\rm During the first stage of cued retrieval, local (featural) dynamical attractors oscillate periodically between a level of high-robustness (HR) and a level of low-robustness (LR); in the second and final stage, robustness is constantly at the lower level. During LR semi-periods, any active feature in any module is stable if the compound association weight onto it is larger than 1; otherwise the module becomes quiescent. During HR semi-periods, any active module will remain active, although it can be forced to switch feature as specified in Law~\ref{law3}. \label{law1}
\end{law}
\begin{law}[\it Activity spread] 
\rm During HR semi-periods, any quiescent module with active neighbours is driven to retrieve a feature onto which at least one of those reproduced by the neighbours has positive association weight. If different neighbours would drive the module into different featural attractors, the module will reactivate a feature randomly chosen among those onto which the compound association weight has the largest value.
\label{law2}
\end{law}
\begin{law}[\it Attractor switch] 
\rm During HR semi-periods, any active module will stay stable if the compound association weight onto one of its other features is not larger than 1 and than the association weight onto the currently active feature; otherwise, the module will switch to the attractor of a feature randomly chosen among those onto which the compound weight has the largest value.
\label{law3}
\end{law}

Randomness is therefore present in the model mostly in quenched form: the architecture of the intermodular connections, the involvement of any module in any pattern, and the specific local feature represented by any module recruited by any global pattern. Dynamical randomness is only present in the choice of feature to be activated in any module during retrieval spread if several feature are equally eligible, be it in the activation of a silent module or in attractor switching between features of an already active module. Although such stochastic effect is relatively small and will be imperceptible in the plots from the simulations, it is in play and is visible in the numerical outputs.

\section{Results}
In this Section, properties and behaviour of the network defined above are studied specifically in regards of cued retrieval and stability of memory reactivation against increasing storage load. 

\subsection{Bounds from architecture}
The modules that are isolated (i.e., without any neighbour) cannot be affected by the dynamics of the rest of the network and, if activated by the cue, will become silent in the first LR semi-period that follows. A similar argument holds for the modules that belong to trees, that is, connected subsets of nodes with no cycles. If a tree is not cued, it cannot be activated through network dynamics. If a tree is cued, activity will remain during the oscillatory stage and possibly spread within it. What happens in the second stage depends on whether the network has LRSS or not, as any feature can only remain stable if there is an association weight larger than 1 onto it: in the nLRSS network,
this can also happen because of pairwise association weights larger than 1, but in the LRSS network it can only happen in the presence of graph cycles \citep{CFM2004}, not in trees. Therefore, a general upper bound to retrieval performance that any realistic dynamics is expected to obey is determined by the fraction of FG modules that are in non-cued trees, including the special case of isolated modules.

The value of the mean coordination number adopted here, which is biologically reasonable, makes the number of modules that are either isolated or in trees negligibly small. However, the subgraph made of the active modules in the pattern to retrieve has mean coordination number equal to about $z t_{1}$, which is significantly smaller than $z$; the trees of this subgraph, that is, the {\it activity isles}, may constitute a non-negligible fraction of the pattern. In fact, it can be calculated that, in large networks, the fraction of FG modules that are not in non-cued trees is given by
\be
G(\varrho) \simeq 1-\sum_{n\ge 1}\frac{(1-\varrho)^{n}}{n!}\left(n z t_{1} \right)^{\sss{n-1}} e^{-n z t_{1}},
\label{noisles}
\ee
where $\varrho$ is the fraction of the pattern to retrieve that is presented as a cue (the modules activated by the cue are randomly chosen among those in the FG of that pattern). It can be shown that $G(\varrho)$ increases monotonically as $t_{1}$ increases, which means that retrieval performance is better for larger $t_{1}$; however, there is also a nontrivial upper bound to the value of $t_{1}$, which is a function of $\tau$ and $z$ \citep{CFM2000}.

In the simulations that follow, the biologically plausible values $z=15$ \citep{Pucak1996,Ichinohe2012} and $\tau=0.1$ will be adopted, given which it can be set $t_{1}=0.25$ (below the upper bound to correlation). With these values, the population of isolated FG trees altogether accounts for an appreciable fraction of the network. This becomes of some relevance when a small cue is presented; for instance, with $\varrho=5\%$, one has that about $1-G(0.05)\simeq 2.5\%$ of the FG of the pattern cannot be retrieved. As an example of Griffiths rare regions \citep{Griffiths}, the activity isles would be almost trivial if it was not for the fact that they are not defined by just the architecture of connections, but also, and mainly, by the pattern of activity that is globally reactivated: the locations of the rare regions vary across the set of patterns, though their total extent stays about the same if the network is large.

\begin{figure}[t]
\center\includegraphics[width=8cm]{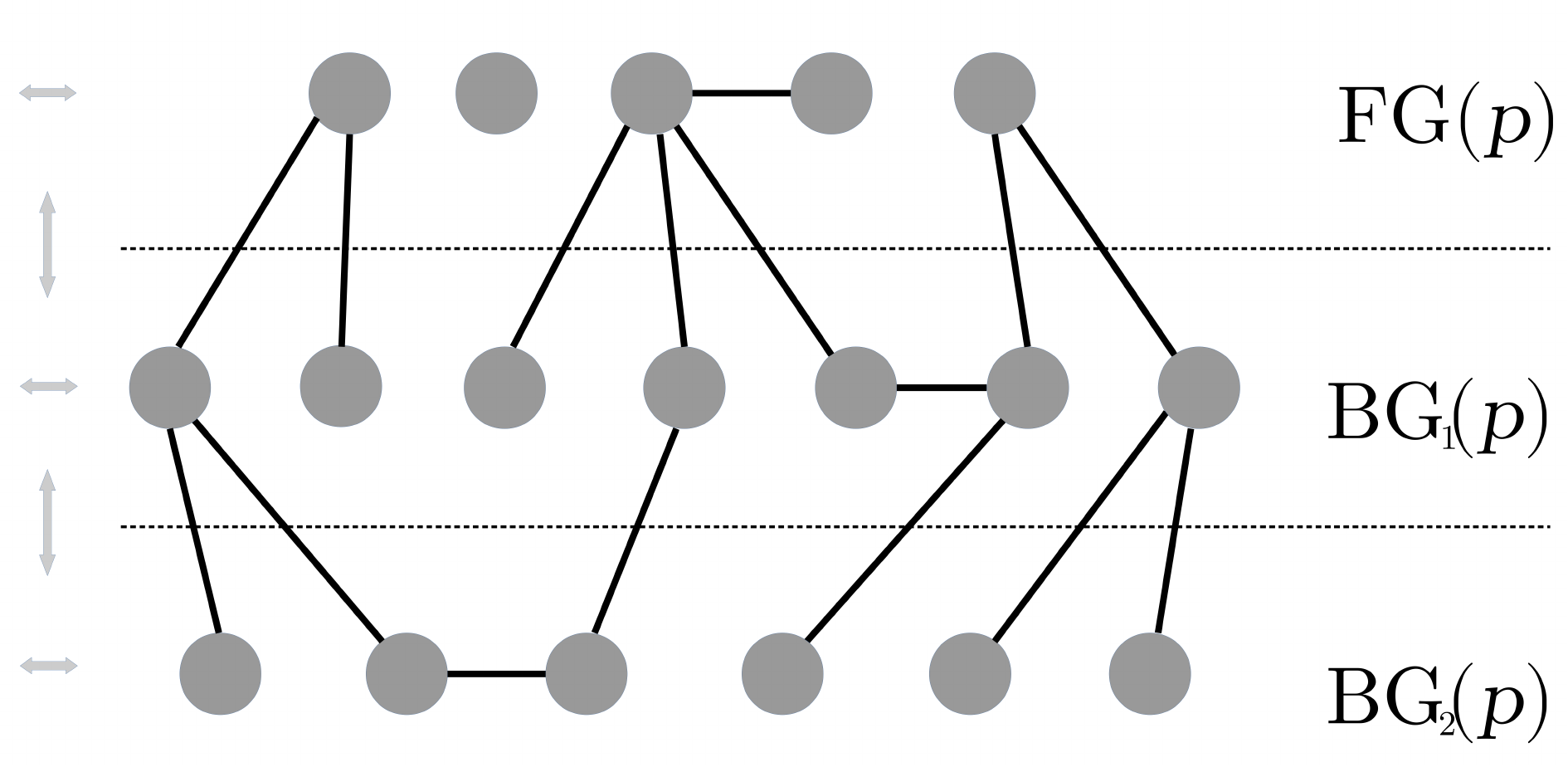} 
\caption{\small{Illustrative diagram of the modular network when pattern $p$ is to be retrieved. From top to bottom: set of modules of the foreground of pattern $p$; set of modules of the background that connect directly to foreground modules; set of background modules that do not connect directly with foreground modules. The shaded arrows on the left show the directions of possible interactions between modules.} \label{cued}}
\end{figure}

\begin{figure*}[t]
\hspace{-.5cm}
\center{\mbox{\includegraphics[height=7cm]{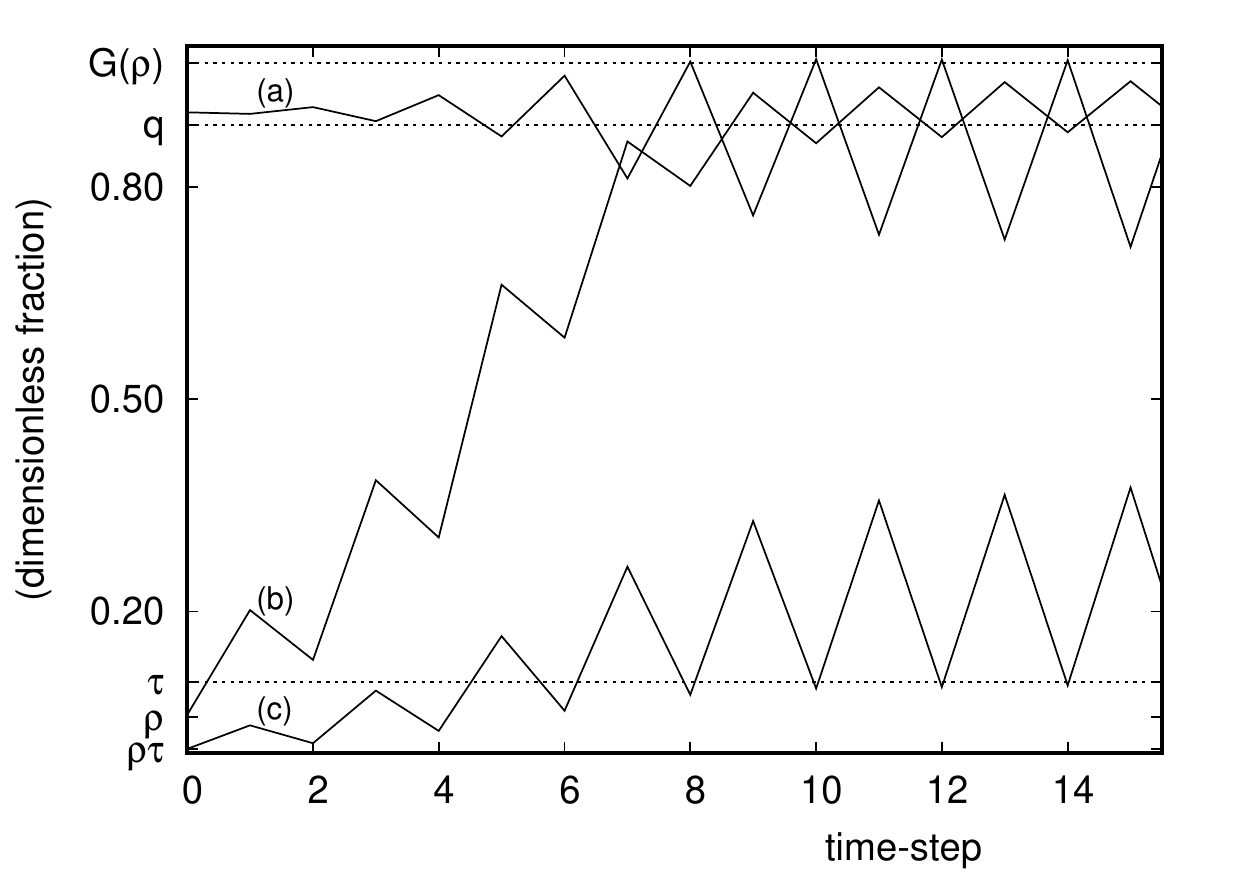}\hspace*{-1.2cm}
\includegraphics[height=7cm]{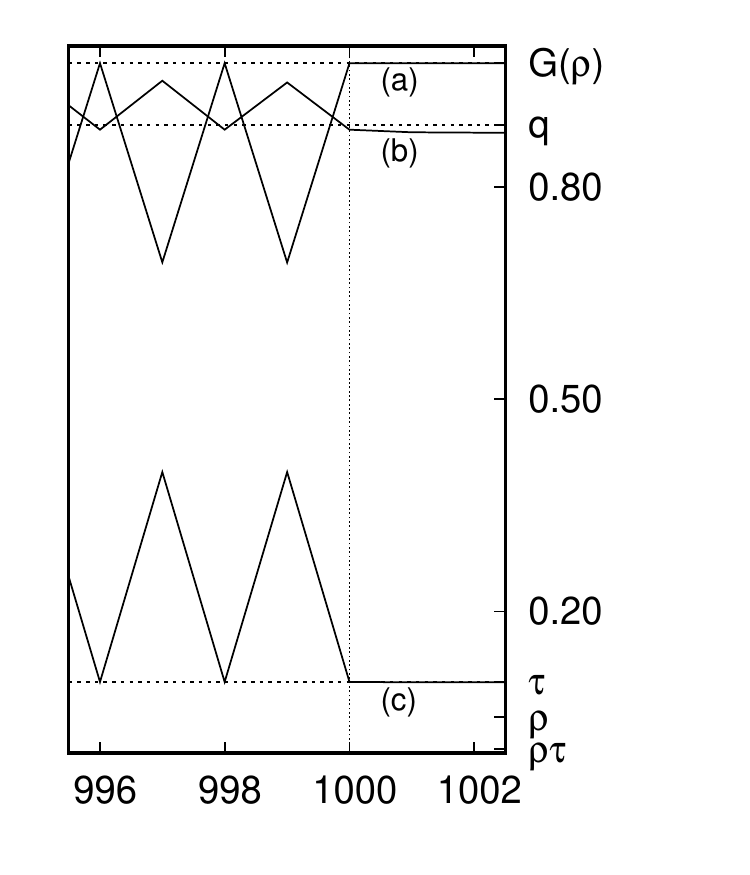}}}
\caption{\small{Cued retrieval process: Output from a typical simulation, with $P=4000$, $F=300$, $\tau=0.1$, $t_{1}=0.25$, $\varrho=5\%$, $M=25,000$, $z=15$. The dashed lines represent the predicted values of, from top to bottom, the maximum achievable retrieval quality $G(\varrho)$ during the first stage, the maximum achievable retrieval quality $q$ during the second stage, and the average activity level $\tau$ of the pattern to retrieve. Continuous lines: (a) Fraction of modules in the respective correct states (quiescence or active feature); (b) Fraction of FG($p$) modules reproducing the respective correct features; (c) Fraction of active modules. The oscillatory process stops at time=1000. \label{retfig}}}
\end{figure*}

In order to estimate the best retrieval performance achievable in the second and final stage of the cued retrieval task, assume that the network is exactly reproducing a memory pattern. The probability $q$ for any FG module to be reproducing a feature onto which the association weight is larger than 1, which would allow for the module to remain in stable activity, is (for large $M$)
\be
q\simeq 1- e^{-z\,t_{1}}-z\,t_{1}\, e^{-z\,t_{1}} \, \left(1- \frac{\tau t_{1}}{F^{2}}\right)^{P-1}
\label{q}
\ee
for the nLRSS network, and
\be
q\simeq 1-(1+z\,t_{1}) \; e^{-z\,t_{1}}
\ee
for the LRSS network. The two formulae differ because in the LRSS case any FG module requires at least two FG neighbours in order to remain in stable activity, while in the nLRSS case one FG neighbour suffices if the latter is reproducing a feature that has an association weight larger than 1 onto the feature in the former; this also explains why the numbers of patterns ($P$) and features ($F$) only appear in the first formula. One can contemplate special local configurations in which modules cannot be driven to stable retrieval activity and which, therefore, lower the upper bound a little further, as for similar cases shown in \can{CFM2004}; however, such minor contributions turn out to be negligible, especially in consideration of the finite-size effects. It may be noticed that $q$ increases if $t_{1}$ increases, which further strengthens the conclusion that pairwise activity correlation is helpful in memory retrieval, besides possibly in the economy of white-matter resources.

\subsection{Cued retrieval}
In Fig.~\ref{cued} a very small portion of the network is pictorially rearranged so to illustrate qualitatively the dynamical process. The modules in the FG of the pattern $p$ to retrieve constitute the set FG($p$) in the top layer. The set BG$_{1}(p)$ of modules that do not belong in the pattern but are connected to FG($p$) modules is in the middle layer. The set BG$_{2}(p)$ of modules that do not belong in the pattern and are not connected to FG($p$) modules is in the bottom layer. The construction could go on iteratively, though with the values of $z$ and $\tau$ adopted here only a tiny fraction of the modules do not belong in any of the layers shown (cf. Appendix A). The arrows on the left indicate the flow of possible spread of activity; as connections are reciprocal, modules in a layer can affect modules in the same layer as well as modules in any adjacent layer. It should be noticed that the layered arrangement depends on the chosen pattern; in different patterns, the modules that constitute the layers are generally different.

At the start of the cued retrieval task, the network is assumed to be quiescent. After a memory pattern to retrieve is chosen, each of the modules that should be active in that pattern is cued by the appropriate feature with probability $\varrho$ and remains silent otherwise. Immediately after cueing, oscillations begin (with a HR semi-period) as determined by the dynamics defined in Section~\ref{dynamics}. In order to verify time-asymptotic behaviour, the oscillations are let to last enough for the network to reach a stationary regime, although this might be unnecessary in real cognitive tasks. Then, the oscillations stop and robustness is kept at its lower level. The first stage of the process serves the purpose of spreading memory retrieval. The second stage is included for the verification of some dynamical properties, though it may be not relevant to actual cognitive processes; in fact, the quality of retrieval is a little higher in the LR semi-periods of the oscillatory stage than in the static final stage.

Figure~\ref{retfig} shows the output of a typical numerical simulation of a cued retrieval task in the nLRSS network (the outputs for the LRSS network are qualitatively identical, with small quantitative differences). The network has $M=25,000$ modules and the fraction of the pattern to retrieve presented as a cue is $\varrho=5\%$ ($z=15$, $P=4000$, $F=300$, $\tau=0.1$, $t_{1}=0.25$). Line (a) is the fraction of modules in the respective correct state, active or quiescent that be; line (b) is the fraction of FG modules in correct features; line (c) is the fraction of active modules (all as functions of time). In the HR semi-periods, all the BG modules that become active contribute to incorrect activity, as do a number of FG modules because of multiplicity of featural associations, while the rest of the newly activated FG modules contribute to correct activity (b). After a few oscillations, at the HR peaks about 95\% of the features are correctly retrieved, quite close to the upper-bound for the network without feature-sharing $G(\varrho)$, Eq.~\ref{noisles}; there is also significant wrong activity, as evidenced by the value of line (c) being much larger than the average activity $\tau$ in the memory pattern, but the set of incorrect features has negligibly small overlap with any memory pattern throughout (not shown). At the LR troughs, the incorrect activity is negligible, its extent being one order of magnitude smaller than that of correct activation, while about 88\% of the correct features are still maintained active, close to the inherent upper-bound $q$ given by Eq.~\ref{q}. (Cf. Appendix A.)

It may be noticed that it only takes about 5 oscillation periods to achieve maximum retrieval; assuming speculatively that the oscillations are related to brain waves in the gamma range, with a frequency of, say, 50~\!Hz, the time needed for maximum memory recall is about 100~\!ms, which seems plausible enough and would fit within a semi-period of a 5~\!Hz theta rhythm, the latter having been often hypothesised by other authors to provide a `clock' for some cognitive processes.

\subsection{Storage capacity}

\begin{figure}[t]
\center\includegraphics[width=8cm]{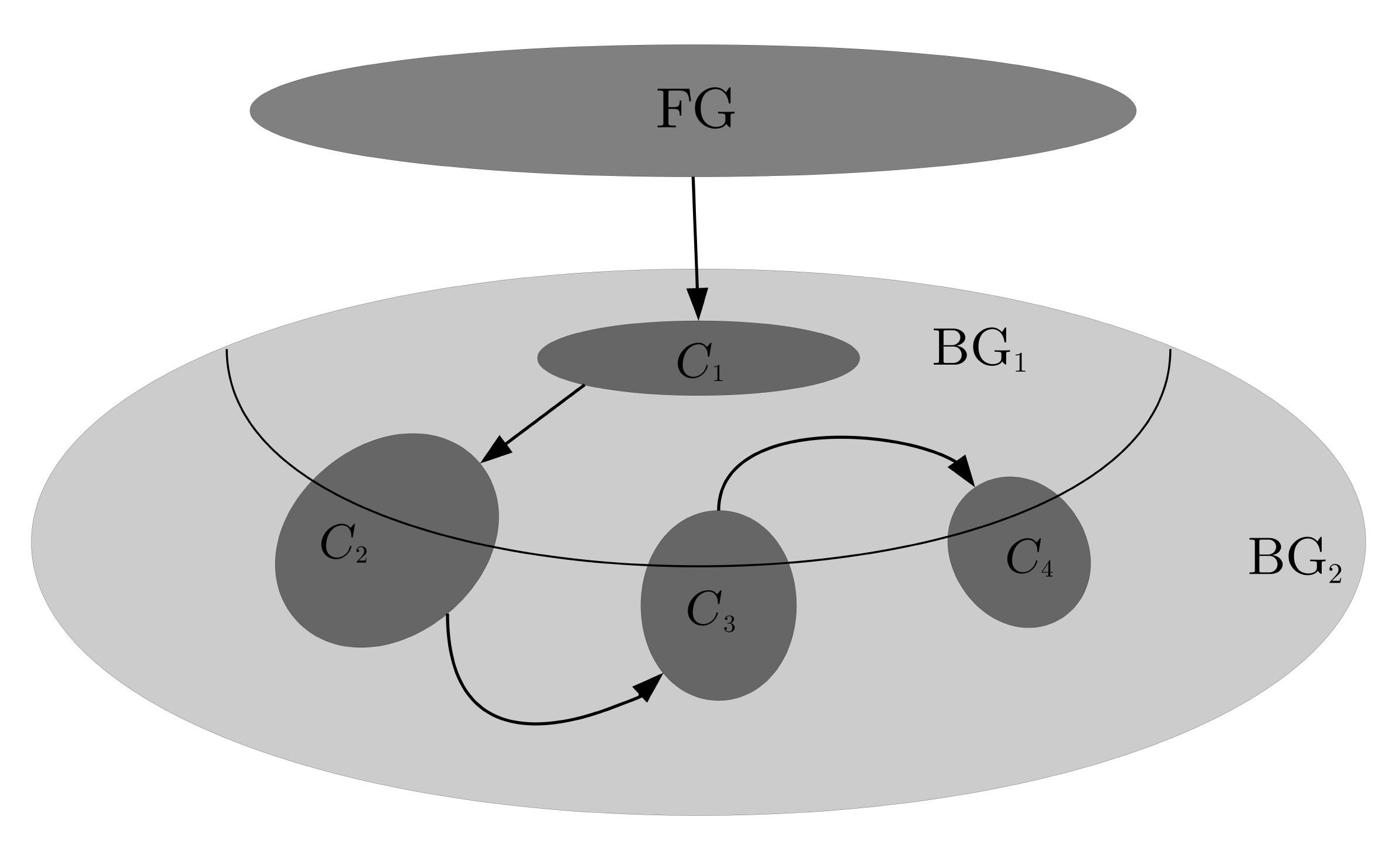}
\caption{\small{Venn diagram illustrating the possible spread of stable incorrect activity. The darker grey areas $C_{1}$, $C_{2}$, $C_{3}$, and $C_{4}$, belonging in the background of the reactivated pattern, represent stable incorrect activity during successive low-robustness semi-periods. The set BG$_{1}$ comprises the BG modules that are directly connected with FG modules, while the modules of BG$_{2}$ connect with modules in BG$_{1}$ but not in FG. \label{venn}}}
\end{figure}

The Venn diagram in Fig.~\ref{venn} illustrates qualitatively the dynamics of activity spread through the network and serves the description of the processes that took place in the simulations that were run to estimate the critical storage-load $P_{c}$. At the top is the set of FG modules, all active in the correct respective features by initial condition; this choice was made because it provides an upper bound, as the likelihood of wrong activity spreading is higher when more modules are active. The BG modules that do not belong in BG$_{1}\cup$~\!BG$_{2}$ are very few and, hence, are here neglected (Appendix A). At the initial step, the FG modules activate BG$_{1}$ modules, some of which, namely $C_{1}$, remain active in the following LR semi-period. Similarly, $C_{1}$ modules then produce stable incorrect activity in $C_{2}$, which contains other BG$_{1}$ modules as well as some BG$_{2}$ modules. Because of recurrent connectivity, $C_{2}$ activates other modules both in BG$_{1}$ and BG$_{2}$. Activity spread continues in the same manner until a stationary state is reached. Below the critical load, the active sets besides FG are very small, while above it they include most of the network, reproducing features that are altogether not coherent with any stored pattern and bringing the mean activity to much larger values than a correct retrieval state would.

\begin{figure}[t]
\center\includegraphics[width=8cm]{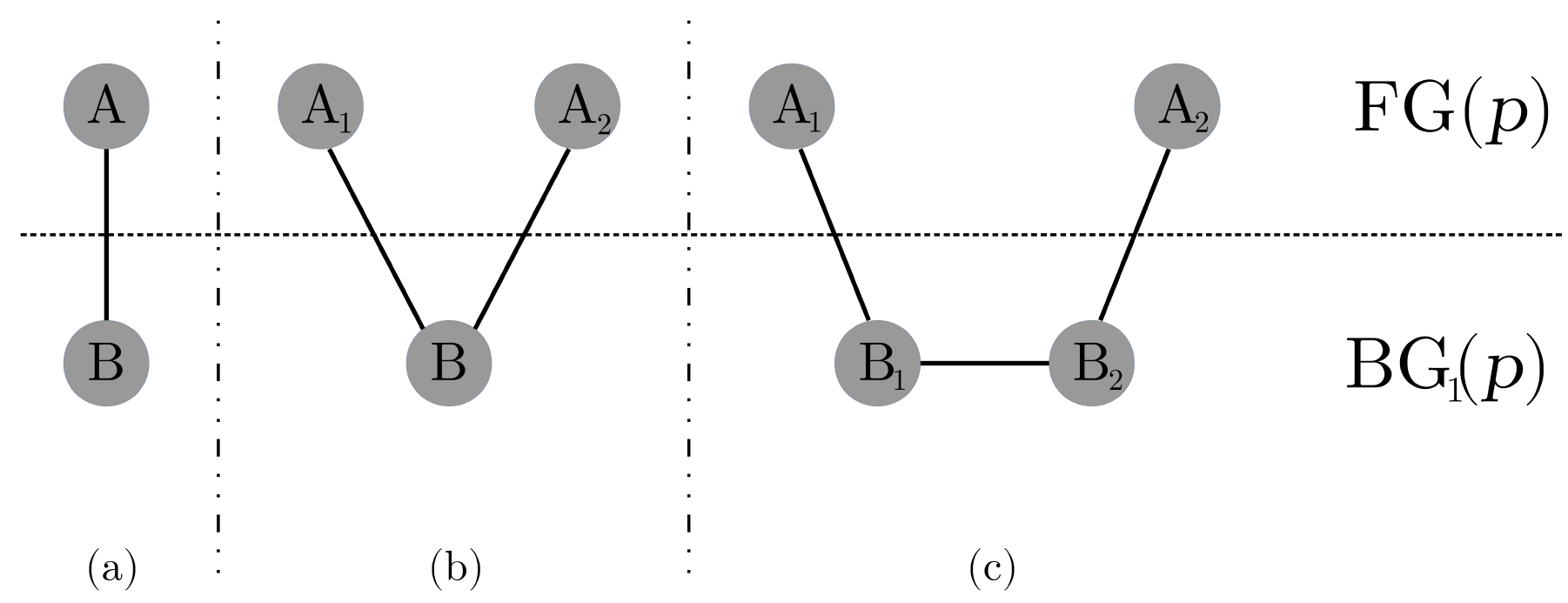} 
\caption{\small{Small subgraphs of special relevance in the qualitative and quantitative analyses of activity spreading.} \label{cases}}
\end{figure}

Figure~\ref{cases} shows the main small subgraphs of relevance in the dynamical spread of activity, which should be considered in isolation for the sake of the following description. Fully connected triplets of modules (3-cliques) are present in negligible number and the corresponding cases are therefore ignored.

Figure~\ref{cases}a shows the case in which one BG module is connected to one only FG module. Module ${\mathbf B}$ is likely to be activated by module ${\mathbf A}$ during the HR semi-period. In the LRSS network, the association weight onto the feature reactivated in ${\mathbf B}$ cannot be larger than 1 and, therefore, in the following (LR) semi-period module ${\mathbf B}$ will become quiescent. In the nLRSS network, module ${\mathbf B}$ will instead stay stable if there is an association weight larger than 1 onto the reactivated feature.

Figure~\ref{cases}b shows the case in which one BG module is connected to two FG modules. If LRSS is present, activity induced in module ${\mathbf B}$ can only survive the LR semi-period if the feature it is reproducing is compatible with both the features reproduced respectively by modules ${\mathbf A}$$_{1}$ and ${\mathbf A}$$_{2}$. In the nLRSS network, it would be also sufficient that the association weight from either of the two FG modules onto the feature in ${\mathbf B}$ be larger than 1.

Figure~\ref{cases}c shows the case in which two BG modules are connected to each other and each to a different FG module. If the respective features they retrieve are compatible with each other, then both ${\mathbf B}$$_{1}$ and ${\mathbf B}$$_{2}$ will stay stable in the following LR semi-period because there is a compound association weight larger than 1 onto each of the features they respectively reproduce. This holds for both the LRSS and the nLRSS network; in the nLRSS network, there is the additional possibility that any feature in the BG modules be kept stable by multiple association with the respective FG neighbour. 

The three main cases described above suggest that the nLRSS network is more prone to spread of incorrect activity, as there are more possibilities for incorrect activation. Consequently, the storage capacity should be expected to be smaller than in the LRSS network.

The numerical study consisted of several batches of simulations on several realisations of the underlying random graph.\footnote{Simulation numbers are not specified because they varied through the study depending on the magnitude of fluctuations. Whenever possible, it was chosen to keep running simulations until the sample standard deviation became stationary. The pseudo-random number generator was adapted from routine {\it run2.c} of \can{NR}.} For each graph, a large set of memory patterns was created. Every simulation batch was composed of several trials, each organised as follows: the network was loaded with a set of memory patterns below the critical value; one of these pattern was then reactivated as initial condition (equivalent to a memory cue with $\varrho=1$); one simulation was run; at the end of the run a predefined number of new memory patterns were added; another simulation was then run; such sequence was repeated until several consecutive runs had showed that the load was above critical storage; then, all memory patterns were replaced and a new trial was started. In any trial, while the number of stored patterns increases towards its critical value, wrong activation gets larger during the HR steps, but it is almost entirely silenced in the LR steps. Extensive spread of stable wrong activity happens suddenly upon storage load reaching the critical value, which makes it easy to measure it, within the limitations due to increment size; however, such critical number depends significantly on the specific set of patterns that is stored because of finite-size effects, which made it necessary to run a considerable number of simulations with different memory sets in order to achieve a reliable estimate.

\begin{figure}[t]
\center\includegraphics[width=8cm]{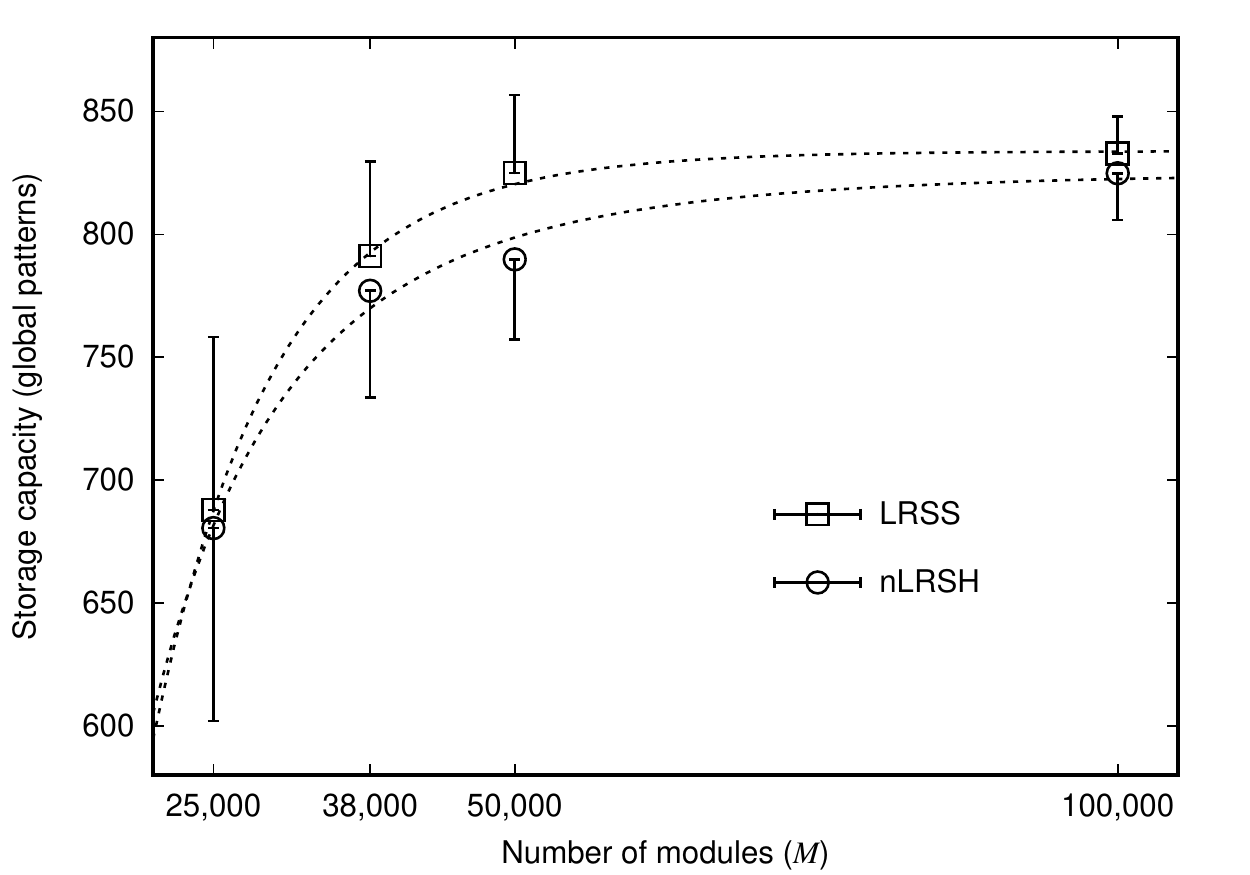}
\caption{\small{Storage capacity of the modular network as a function of the number $M$ of modules, the number of features per module being constant ($F=100$). The segment ending in each data point is as long as the sample standard deviation (to avoid graphics overlap, it is shown only above or below the corresponding point). \label{PvsM}}}
\end{figure}

\begin{figure*}[t]
\vspace*{-.3cm}\hspace*{-0cm}
\center{\includegraphics[width=14cm]{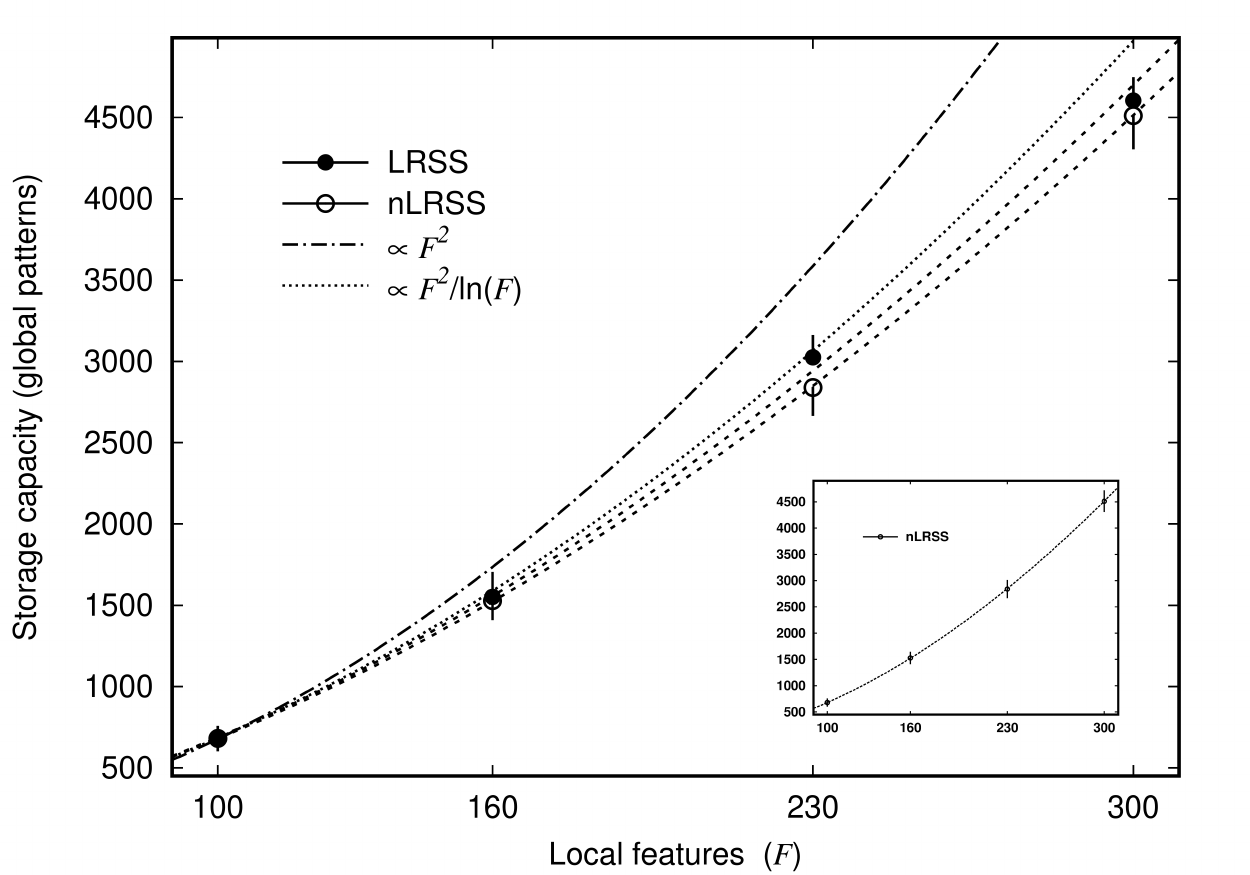}
\caption{\small{Storage capacity of the modular network as a function of the number $F$ of features stored in each module, the number of modules being constant ($M=25,000$). The data points as well as their respective best-fit curves are shown for the LRSS network and for the nLRSS network. The segment ending in each data point is as long as the sample standard deviation (to avoid graphics overlap, it is shown only above or below the corresponding point). The curve for a purely quadratic function as well as its correction by a logarithmic denominator are also shown for comparison, both fitted to the two almost coincident data-points for $F=100$. For clarity, the inset re-presents the data points and best-fit curve for the nLRSS network alone. \label{PvsF}}}}
\end{figure*}

Eventually, mean and sample standard deviation of the critical values were calculated. One large set of batches was run on graphs of different sizes while keeping the number of features constant ($F=100$), the respective critical storage-loads being plotted in Fig.~\ref{PvsM}; it may be noticed that the standard deviation is smaller for larger $M$ (in fact, the variance is about inversely proportional to $M$), which supports the idea that very large networks will behave almost identically (self-averaging). A second large set of batches was run with different numbers of local features while the size of the graphs was kept constant ($M=25,000$), the respective critical storage-loads being plotted in Fig~\ref{PvsF}.

Figure~\ref{PvsM} presents a crossover between the two capacity curves at a value of $M$ below those simulated. It is possible that at smaller network sizes higher order corrections for finite-size effects be required; the crossover, anyway, disappears for more realistic values of $F$ (e.g., $F\geq 1000$).

The demand on computer RAM scales approximately as $M\cdot F^{2}$, making it unfeasible to simulate the larger networks using the larger numbers of features. A heuristic analysis (Appendix A) guided the search for a functional form of the (average) critical storage-load that agree with both the plots ($F$,~$P_{c}$) and ($M$,~$P_{c}$). The numerical values of three of the four undetermined parameters were estimated by best-fit of the ($M$,~$P_{c}$) data plot (Fig.~\ref{PvsM}, $F\equiv 100$), which was derived from a richer data set.\footnote{As the standard deviation is smaller for larger $M$ and is due to inherent finite-size effects, it was chosen to use a simple minimisation of sum of square errors in order to avoid underweighting the mean values at lower $M$.} Finally, the value of the remaining parameter was obtained from the ($F$,~$P_{c}$) data plot (Fig.~\ref{PvsF}). This procedure led eventually to
\be
\left\{
\ba{l}
P_{c} \simeq  \, a \, h(F,M) \frac{\dsty{F^{2}}}{\dsty{F^{h(F,M)}}-1} \\ \\
h(F,M) \doteq \beta \dsty \left(\ln F\right)^{\frac{1+2 \mu}{4}} 
e^{-\nu M^{2(2-\mu)/5}},
\ea
\right.
\label{Pc}
\ee
where: $\mu\simeq 0.00$, $a\simeq 0.384$, $\beta\simeq 1.5$, $\nu\simeq 0.001$ in the LRSS network, and $\mu\simeq 1.00$, $a\simeq 0.380$, $\beta\simeq 6.0$, $\nu\simeq 0.095$ in the nLRSS network. The parameters $a, \beta, \nu$ depend on the network parameters $z$, $\tau$, $t_{1}$, though likely in different ways for the two types of network; alas, a numerical study of such dependencies was prohibitive in terms of computing resources. The values of $\mu$, however, seem more likely to reflect the combinatorial dissimilarity in the dynamics, though one cannot disregard the fact that the simulation data have significant uncertainties. It can be noticed straightaway that, for large $M$, one has
\be
P_{c} \propto \frac{F^{2}}{\ln F}.
\ee

For a nLRSS network with $M=25,000$ and $F=20,000$, both fairly realistic sizes, Eq.~\ref{Pc} gives $P_{c}\sim~\!\!10^{7}$, which seems plausible. It may also be observed that the average number of features in any module that are associated to any given feature in any adjacent module through a set of stored patterns is
\be
\left[1-\left(1-\frac{\tau t_{1}}{F^{2}}\right)^{P}\right] F,
\label{multi}
\ee
whose value with a storage load just below capacity is about 12.5, which appears to confirm that local ambiguity of associations, consequence of feature-sharing, is indeed present and relevant. As an aside, it may be noticed that, for $F=20,000$, the storage capacity of a network where finite-size effects are no longer relevant ($M>100,000$) would be predicted by Eq.~\ref{Pc} to be about 10 times that of a network with $M=15,000$ (extrapolations to lower values of $M$ are likely unreliable).

\section{Discussion}
\subsection{Experimental background}
The first general model of a modular organisation of the neocortex was probably the one proposed by \can{Braitenberg1978}. The author based his idea mainly on observations about cytoarchitectonic properties: (1) the pyramidal neurons seem to constitute the computational `skeleton' of the cortex, with the other types of neurons mainly playing a supporting role, (2) the targets of short and, respectively, long range axonal projections are mostly in different layers of the cortical sheet, hinting at some degree of differentiation between local and global processing, and (3) the number of long-range (inter-areal) projections seems far too small to support the theory of an approximately homogeneous architecture. He suggested that the neurons of the cortex are grouped so to form a large number of modules: recurrent connectivity inside any module is dense, while connections between modules are sparse but still enough for the modules to work as a whole. This anatomy-based hypothesis is in fact compatible with the then already available evidence of columnar structures and functional modularity in sensory cortices \citep{Mountcastle1957,HubelWiesel1977}. Later, further support for functional modularity, importantly including local reverberation, arrived \citep{FusterJervey,Miyashita1988,MiyashitaChang}, as did more detailed connectivity studies \citep{GoldmanRakic,Tommerdahl,Pucak1996,Ichinohe2012}. Because of experimental constraints, the observed cortical patches in each experiment are few and quite limited in size; therefore, long-range projections are likely undercounted, thus leading to connectivity statistics that might be significantly biased towards a greater dependence of connection probability on distance, some indication of this possibility being apparent, for instance, in the data analysis of \can{Ichinohe2012}. Convincing evidence of large scale recurrent processing in Primates also appeared \citep{Hasegawa,Tomita,Naya,Miyashita2011,Miyashita2013,Miyashita2015,Miyashita2017}, which showed emphatically that recurrent memory processing is not limited to local ensembles and involves long-range interactions between plausibly a large number of them, with laminar functional differentiation also being evident. Notwithstanding the correspondence between functional and anatomical columns being debated \citep{Jones2000,Rakic2007,Lent2008,Kaas2012}, it is quite clear that functional ensembles are geometrically close to be `columns' in only some cortical areas, while being more stripey in others \citep{Pucak1996}. It may be speculated that functional modules derive from coalescence of the phylogenetic mini- or micro-columns, that are known to appear during embryonal development of the cortex.

\subsection{Theoretical background} \label{theories}
The model of \can{OKT} implemented the modular organisation proposed by \can{Braitenberg1978} quite faithfully. Each module is a Hebbian associative network of the kind studied previously by \can{Treves1990} as a more biologically-realistic development of earlier models by other authors \citep{Little1974,Hopfield1982,Amit1985,TsodyksFeigelman}. Highly dense recurrent connectivity allows each module to store a large number of local features, while the long-range projections from any neuron of any module are distributed randomly across the neurons of the other modules. If, adhering to anatomical data \citep{BraitenbergSchuz}, the number of intramodular afferents per neuron is made similar to that of extramodular ones, the long-range connectivity results to be much more dilute than the local one and, hence, any memory pattern may be considered as made of a specific combination of local features. By employing machinery from equilibrium statistical mechanics in order to study the existence and stability of global memory retrieval states, \can{OKT} found that the landscape of (metastable) network states was largely occupied by `memory-glass', a phase in which modules reproduce respective features locally correct but globally incoherent, so that the network as a whole is, so to speak, in a meaningless state. The global retrieval states are only slightly more robust to noise than memory glass, thus requiring a very fine tuning of noise level that led the authors to question the biological plausibility of a theory of the neocortex as a large complex of interacting autoassociators. 

It was later shown \citep{FMT,CFM2000x} that including the almost obvious requirement that memory patterns be more realistically sparse, so that only a small fraction of modules are involved in any pattern instead of all of them, only improves the network ability marginally. Making the axonal projections from any module be distributed across a relatively small number of the other modules instead of (statistically) across all of them does not lead to any improvement. However, if the requirement is further added that the activities of any connected pair of modules be correlated across the set of learned patterns, then the proper retrieval states gain a large advantage in robustness with respect to the memory-glass states.

\noindent \quad A number of questions remained open, of course. Among them, one was about the retrieval dynamics: the study was addressing the existence of metastable states, but not if and how they may have been reached with biologically realistic dynamics and initial conditions. Another was the estimate of the storage capacity of a multimodular autoassociator: indeed, the signal-to-noise analysis by \can{FMT} was local and could not take into account large-scale cooperative and combinatorial effects.

The first work to address the dynamics of biologically realistic modular autoassociators was probably the one by \can{Parga}. The authors used a firing-rate neuron model to study a network made of three modules, with reciprocal Hebbian synapses between one module and the other two, in the idea that the central module could model a cortical area at a higher level in the information processing hierarchy than the others. They found that the network can work in several regimes: depending mainly on the value of the parameter that tunes the relative strength of intermodular and intramodular neuronal interactions, the modules can behave with different degree of independence and the retrieval activity can spread more or less easily between them. Their model is capable of reproducing qualitatively well known misperception phenomena due to contradictory inputs from two modalities. However, limiting the network to just three modules, while making it more amenable to an analytical approach and yielding enlightening results, also limited the scope their conclusions could have on networks of several recurrently connected modules.

\noindent \quad The dynamics of the spread of neuronal activity through a network of many modules was studied in \can{CFM2004}, mainly concerning cued memory retrieval. It was shown that the spread of incorrect activity caused by the local ambiguity of associations (pairwise featural associations are not univocal), which would lead to a meaningless patchwork of pieces from many memory patterns alike memory glass, is mostly prevented if local attractor robustness to noise undergoes periodical oscillations, with a modular dynamics accordingly defined. Differently from the present work, the requirement that any active feature be able to stand destabilising noise was that it has at least two supporting neighbours, having not included, as a first approximation, the role of multiple pairwise association weights. While the laws of the dynamics were defined ad hoc, they turned out to be very compatible with the behaviour of the network of \can{Parga}.

In \can{CFM2004}, stability of local feature activity is estimated by neuronal signal-to-noise analysis, after which the modular network is treated at a coarse-grained level of multi-state dynamical units, with explicit inhibition being within the intramodular neuronal network only. Having chosen to work directly with modules as multi-state units, \can{Treves2005} adapted the statistical-mechanical model of Potts neural network to study properties of multimodular autoassociators. He extended results previously obtained by \can{Kanter1988} in order to analytically estimate the storage capacity of the modular model at equilibrium, and, after having endowed the modules with a form of adaptation that weakens in time the global attractor in which the network stays, studied by numerical simulations the possibility for the network to travel indefinitely through the set of learned patterns. The synaptic `weights', as in \can{Kanter1988}, are in fact matrices and are allowed to induce both excitation and inhibition of specific features as a consequence of being a generalisation of the covariance rule of \can{Hopfield1982}, in marked contrast with the model presented here, wherein feature-specific intermodular inhibition is not allowed. Whichever of the two contrasting assumptions will result to be more realistic, it is noteworthy that the functional dependence of storage capacity on (large) number of local features is, in the large-network limit, the same for the two models. The precision in the calculation of the storage capacity in \can{Treves2005}, however, comes at the price of bypassing the problems of local ambiguity of associations and of memory glass because of a stochastic dynamics that does not account for robust persistence of local reverberations.

Other models of multimodular networks have subsequently been proposed, though a (brief) comparison will only be made with some of the closest ones. \can{Lansner2013} studied an attractor network in which minicolumns are grouped into hypercolumns and connectivity is patchy. They found that it is possible to include representation sparseness and synaptic pruning in their model to make it more realistic without compromising storage capacity, which remains plausibly large; however, only one feature is there associated to each unit, while synaptic pruning does not work as a homeostatic mechanism. With a view at representing concept categorisation, \can{Brunel2016} used binary synaptic weights and sparse distribution of patterns of activity across modules of binary neurons to reproduce semantic relations, achieving plausible storage capacity and ability to complete autoassociative tasks (pattern correction or completion) both in individual modules and in the network as a whole. A semantic correlational structure arises in their model as a consequence of how pattern representations are distributed over subsets of modules, conspicuous differences with the present work being though that their architectural intermodular connections are determined by the learning process and that patterns do not share features. \can{Treves2018} made a Potts neural network learn sets of memory patterns with an explicit hierarchical organisation that departs from the simple tree-branching and is compatible with the correlational structure in a data set of nouns, which in terms of storage capacity only costs a minor reduction. While all these models present characteristics that are to a large extent compatible with, or complementary to, those presented here, the notable conceptual and mathematical differences in a priori statistics of the memory patterns or modular dynamics make a much deeper comparison difficult.

As in all the models of memory storage and retrieval cited above, the present model concerns memory performance after memory storage, for only the outcome of the learning phase is taken into account. A learning process in a modular associator with small numbers of `areas' and simpler pattern statistics and connectivity was studied by \can{Papadimitriou2020} in a brain-inspired computer-theoretic perspective which employs binary neurons, a k-winners-take-all algorithm, a simplified Hebbian learning rule, quenched randomness of inter-neuron connections, and weight normalisation throughout learning. Their simulations also verified that their modular network, after learning, is capable of performing basic autoassociative tasks. The most obvious differences with the present model are in the architecture of network and connections and the small number of modules, which imply that representations are highly distributed at the neuronal level but not at the featural level. A possibly more important difference, however, is that modular features are not allowed to be shared across patterns; including feature-sharing would be essential for it to constitute the learning stage of the model presented here.

\subsection{Summary and conclusions}
The dynamics of the network was defined in such a way as to hinder the activation of features that do not belong in the pattern to retrieve, as per the memory cue, but that may be elicited by correctly activated modules due to local ambiguity of association. To this end, the robustness of local attractors was made oscillatory, so that any feature in any given module would survive a low-robustness step only if it had support of at least two associations, thus leveraging on combinatorial unlikelihood.

While no attempt was made to justify the existence of the oscillatory mechanism on experimental evidence, there is no shortage of results that lend themselves to educated conjectures in support of the hypothesis, such as modulation of attractor robustness to noise \citep{Durstewitz2,BrunelWang} and long-range synchronised oscillations \citep{Gray1989,Wang2010}. A speculative relation with known cortical rhythms yielded a plausible timescale for the memory retrieval process.

Cued retrieval is achieved with high quality, almost saturating two upper bounds that are determined by the statistical properties of underlying graph and patterns of activation and that were calculated analytically: the first bound is mostly due to activity isles, which may reflect inability to retrieve small fragments of memory in real cognitive tasks; the second bound is mainly relevant to the stability of the retrieval state. Pairwise correlation of activation, which may have semantic significance, plays a worthwhile role: the quality of retrieval increases if activity correlation between adjacent modules is increased.

Extensive numerical simulations and a heuristic mathematical analysis were carried out to quantify the storage capacity of the network as a function of the number $F$ of features per module and of the number $M$ of modules. In fact, finite-size effects are evident if $M$ is of orders of magnitude that are quite realistic and, therefore, it is meaningful to take them into account. The analytical formula contains few undetermined parameters whose numerical values were estimated from simulation data, resulting in a good fit with the data plots. The fitted curves also appear to confirm that local ambiguity of association is not a negligible phenomenon, as extrapolation to realistically large values of $F$ shows that multiple pairwise associations can be numerous. In the asymptotic approximation of large $M$, the critical storage-load becomes $P_{c}\propto~F^{2}/\ln(F)$, functionally identical to that found in the analytical derivation for extensive Potts neural networks with large $F$ at `thermodynamic' equilibrium \citep{Kropff,Treves2005}; this suggests that the functional form of the dependence of the storage capacity on $F$ in large networks is fundamentally of combinatorial origin. Specifics of, for instance, synaptic weight representation, code sparseness and connectivity are expected to only affect the values of three of the four undetermined parameters, in a way that is not yet analytically explicit in the mathematics of the present model.

The actual storage of all the local features and their associations through a set of memory patterns allowed for investigating the effect of homeostatic synaptic scaling on storage capacity and retrieval abilities. It has been speculated since the results of very early mathematical models of autoassociative memory that some process should exist that could normalise, maybe during sleep, the synaptic weights of a Hebbian network in order to keep it in a working window \citep{Crick1983}. It was also shown in a modular network dynamical model that some mechanism is required to normalise the synaptic weights within each module: because intramodular synaptic-weight increments are of the same sign much more often than those between neurons of different modules, intramodular signals may exceed the extramodular input to the point that multimodular cooperation loses relevance \citep{CFM2004}. Experimental evidence shows that synaptic homeostasis does take place \citep{Delvendahl2019,Turrigiano2008,Tononi}, although it is not known whether a presumably much more complex associative functional scaling of long-range synapses exist. In the present model, the effects of long-range synaptic scaling are found to only yield a minor increase of storage capacity and to be unimportant during memory retrieval. It seems therefore unlikely that the potential advantage of long-range-coordinated synaptic scaling could constitute enough of an evolutionary drive to lead to the appearance of as a complex mechanism as it would require.

The hypothesis that memory patterns involve features of interconnected modules in several combinations and that the activation of one feature in a module may induce activation of one out of several possible features in a connected module is in principle testable by electrophysiological recordings in behaving primates. Also feasible seems testing the prediction that the spread of retrieval activity from active modules to quiescent ones follows oscillations determined by a neuron-unspecific pulsating modulation (e.g., dopaminergic). Testing the prediction of differential homeostasis of extramodular and intramodular associative synapse might also be feasible, albeit likely very difficult, helped possibly by long-range and short-range synapses being mostly segregated respectively into different layers of the cortical sheet.

\subsection{Research outlook}
The presented work largely achieved its objectives, as the dynamics adopted is biologically realistic, wrong activation is mostly suppressed, the quality of retrieval is close to the ideal upper-bound, and the storage capacity is realistically large and compatible with that found in similar models by other approaches. The next step in order of importance should probably be to make the model more realistic by allowing for some wrong activity to randomly survive the destabilising semi-period but not enough to cause robust spread of further wrong activity. Preliminary results suggest that this objective may be fairly easily achieved by introducing global inhibition whose intensity is a monotonic, nonlinear function of the number of active modules. Such modification is compatible with known neuromodulatory mechanisms acting on the neocortex as well as metacognitive theories (since at least the work by \can{Koriat1993}). Neuronal signal-to-noise analysis showed \citep{CFM2004} that active modules should have higher mean firing-rates than quiescent ones, which can be instrumental to regulating activation by inhibitory feedback loop. The preliminary results also indicate that the centralised inhibitory control may provide the network with the ability to correct errors present in memory cues. However, assuming that the model should just suppress all the fragments from several memory patterns that make up the cue but the largest one of them should not be a foregone conclusion: when and how the network function switches between pattern completion, error correction and segmentation, likely under the influence of neuromodulators, is still very much an open question. Similarly, the separation between learning and retrieval stages in association cortex is also a difficult issue, more than, for instance, in the hippocampal memory system, for semantic learning relies on the extraction of salient information across episodes.

It should be emphasised that the connections between modules in the present model are not determined by learning, but, rather, learning is thought to adapt to a pre-existing architecture. It is assumed that features that statistically co-occur more often than chance tend to be stored in adjacent modules, rather than edges between modules being determined by learning, the latter seeming not very plausible for white-matter connections. This in itself may already have semantic relevance, whereby semantic `categories' should be intended as much finer relatives of those commonly found in the literature (e.g., `living things', `tools', `round objects', etc.). Furthermore, it may be conjectured that, while long-range axonal bundles between modules may be at least in part built randomly, the eventual quenched variability of number of neighbours across modules underlie a spontaneous emergence of levels of semantic hierarchy. Indeed, a different number of neighbours may lead to a different degree of semantic value because the features of a module that has more neighbours are subject to larger associative support or to more numerous, different driving inputs. The dynamical approach of the present work seems to be suited to investigating such possibility. 

Even though the emergence of semantic features from a tabula rasa is likely a complex process of statistical learning, a sufficiently large number of features, once in place, may provide `dimensions' onto which new information may be projected, therefore allowing for faster storage while also eluding or lessening memory interference or forgetting (possibly catastrophic); indeed, \cite{Flesch2018} found that the pre-existence of categorical dimensions facilitated statistical learning of new information.

The model is not very amenable to mathematical analysis. Some of the problems of probabilistic combinatorics it presents are nontrivial and possibly of some interest in themselves. Nevertheless, enough progress is conceivably not beyond reach to provide better insight into the dependence of performance on some experimentally measurable quantities and, therefore, endow the model with greater explanatory and predictive power.

\section*{Computer code}
All the information required to replicate the numerical simulations is contained in the article. Original {\it C} code may be obtained from the author upon reasoned request.

\section*{Declaration of competing interest}
No competing interests.

\section*{Funding}
This research did not receive any financial support of any kind.

\section*{Acknowledgements}
Thoughtful questions by an anonymous reviewer led to some significant improvements in the exposition of the work and of its relevance in context.

\appendix
\section{Probabilistic-combinatorial analysis}

A summary of the reasoning and analysis that yielded the formulae presented in the main Sections is here reported. More details will be included in a dedicated, more mathematical paper.

Because of the activity-structure correlation in Eqs.~\ref{scheme}, the subgraphs of, respectively, FG and BG modules have mean coordination numbers that differ from $z$. By means of Bayesian inversion, for large $M$ one obtains that the probability for FG modules to be adjacent is
\be
\mathbb{P}\left({\mathbf A}_{1}\lra{\mathbf A}_{2} \vert {\mathbf A}_{1}, {\mathbf A}_{2}\in {\rm FG}\right) \simeq \frac{z\,t_{1}}{\tau M} ,
\ee
for BG modules it is
\be
\mathbb{P}\left({\mathbf B}_{1}\lra{\mathbf B}_{2} \vert {\mathbf B}_{1}, {\mathbf B}_{2}\in {\rm BG}\right) \simeq \frac{1- t_{0}}{1-\tau}\frac{z}{M} ,
\ee
which is also valid for BG$_{1}$ modules because the events of being in BG and being adjacent to FG are independent, and for two modules in respectively FG and BG it is
\be
\mathbb{P}\left({\mathbf B}\lra{\mathbf A} \vert {\mathbf A}\in {\rm FG},\ {\mathbf B}\in {\rm BG}\right) \simeq \frac{z\,t_{0}}{\tau M} .
\ee
As to be expected, being $t_{1}>\tau$ and $\tau<1/2$ by assumption, the connectivity within FG is larger than the connectivity within BG, which, in turn, is larger than the connectivity between FG and BG (cf. Eq.~\ref{consist}). 

Keeping only the leading order in $M$ as well of binomial averages, one obtains that
\be
{\mathbb E} \vert {\rm BG}_{1} \vert \simeq \left(1 -\tau \right) \left(1-e^{-z t_{0}}\right) M 
\ee
and
\be
{\mathbb E} \vert {\rm BG}_{2} \vert \simeq \left(1 -\tau \right) e^{-z t_{0}} \left[1-e^{-(1-t_{0}) z(1-e^{-z t_{0}})}\right] M ,
\ee
the statistical fluctuations being of relative order $1/\sqrt{M}$. The same calculations can also be carried out for sets further in the sequence; however, inserting the values of the parameters of the networks, it results that at most a few modules do not belong in FG~$\cup$~BG$_{1}\cup$~BG$_{2}$, which union is therefore the only set of modules taken into consideration.

At the first HR time-step of the cued retrieval process, because all FG modules were active in at least one of the patterns during learning, that is, at least in the pattern to retrieve, none of the non-cued FG modules that neighbour cued ones will remain silent; some of the non-cued FG modules may be driven to retrieve incorrect features as a consequence of feature-sharing. For large $M$, the cued FG modules and their FG neighbours at the first HR peak will account for a fraction of the network about equal to
\be
\tau \varrho + \tau \left(1 - \varrho \right) \left(1-e^{-\varrho z t_{1}}\right) , \label{firstpeak1} 
\ee
which therefore becomes active. The modules in BG$_{1}$ that are driven to necessarily incorrect features account for a fraction of the network about equal to
\be
\left(1 - \tau \right) \left[1-e^{-\varrho z t_{0}} 
\exp\left\{\varrho z t_{0} \left(1-\frac{\tau t_{1}}{F}\right)^{P}\right\}\right] , \label{firstpeak2} 
\ee
that also takes into account the possibility for any BG$_{1}$ module to have several FG neighbours. The sum of \ref{firstpeak1} and \ref{firstpeak2} gives the fraction of active modules at the first HR peak. This value agrees very well with those from the simulations (but for minor statistical fluctuations), which provides a simple first check on the validity of the assumptions in the network of the chosen size. A finite fraction of the modules so activated will stay stable in the following LR semi-period; then, the reasoning can be iterated starting from a larger number of active modules, albeit some of them may now be reproducing wrong features. How many modules remain active in LR troughs depends on whether the network is of the LRSS type or of the nLRSS type, which is also relevant in the evaluation of the spread of incorrect activity.

In order to estimate the network storage capacity, one has to count the number of modules that are retrieving incorrect features and are stable during the LR semi-period; if it is large, possibly percolating the network, the retrieval performance is corrupted. The main local configurations to be considered to this end are the small subgraphs in Fig.~\ref{cases}. For the following part, it is convenient to define $\alpha~\equiv~\alpha(M,F)$ such that
\be
P = \frac{F^{2}}{\tau t_{1} \alpha} . 
\ee
Consider first the LRSS network. The case of Fig.~\ref{cases}a, where module ${\mathbf A}$ reproduces feature $f$ and module ${\mathbf B}$ reproduces feature $g$,  is not relevant because there cannot be an association weight larger than 1 onto feature $g$. In the case of Fig.~\ref{cases}b, where modules ${\mathbf A}_{1}$ and ${\mathbf A}_{2}$ reproduce respectively features $f_{1}$ and $f_{2}$, the probability for at least one feature $g$ to exist in module ${\mathbf B}$ onto which the compound association weight be larger than 1 is equal to about
\be
r_{0} \simeq 1-\left(1-\frac{\tau t_{1}^{2}}{F^{2}}\right)^{P} \simeq 1 - e^{-t_{1}/\alpha} .
\ee
The case of Fig.~\ref{cases}c is a little more complicated, for pairs of BG$_{1}$ modules cannot be taken in isolation and, importantly, there may be associations built between the two BG$_{1}$ modules by patterns that only involve one of the two FG modules in question or neither. First, assuming again that modules ${\mathbf A_{1}}$ and ${\mathbf A_{2}}$ reproduce respectively features $f_{1}$ and $f_{2}$, one calculates the average number of features of ${\mathbf B_{1}}$ that are associated with $f_{1}$:
\be
N({\mathbf B_{1}},f_{1}) \simeq \left[1-\left(1-\frac{\tau t_{1}}{F^{2}}\right)^{P}\right] F \simeq \left(1-e^{-1/\alpha}\right) F . 
\ee
The same result, obviously, holds for $N({\mathbf B_{2}},f_{2})$. Then, the probability for any pattern to exist that involve simultaneously one among the $N({\mathbf B_{1}},f_{1})$ features and one among the $N({\mathbf B_{2}},f_{2})$ results to be
\be
r_{1} \simeq 1-\left[1-\tau t_{1}\left(1-e^{-1/\alpha} \right)^{2}\right]^{P} . 
\label{r1}
\ee
With the value of $z$ adopted in this model, a large majority of the BG modules are in fact in BG$_{1}$. All such modules already have support by at least one FG module, which implies that their activity is stable if each of them has support from at least another BG module. Therefore, to evaluate the percolation threshold for incorrect activity, one can consider BG$_{1}$ as a new random graph with edge probability proportional to $r_{1}z/M$; as the number of pairs in BG is proportional to $M^{2}$, one has that, in order to hamper spread of incorrect activity, it must be $r_{1}~<~n/M$ for some finite $n$. This implies that, for any fixed $M$, $\alpha$ must diverge in the $F~\to~\infty$ limit, which would also make $r_{0}$ vanishingly small.

In the nLRSS network, the case of Fig.~\ref{cases}a becomes important. Calling $f$ the feature reproduced by the FG module ${\mathbf A}$ and $g$ the generic feature in ${\mathbf B}$, define the random variables
\be
X_{g}\doteq{\Big\vert} {\Big\{}{\rm pattern}\ p\in [P-1] \ {\Big\vert} \ g \lra f\ {\rm in}\ p {\Big\}}{\Big\vert} , 
\ee
and
\be
X\doteq \max_{g\in [F]} X_{g} . 
\ee
The probability for ${\mathbf A}$ to elicit stable incorrect activity in ${\mathbf B}$ is, for large $F$,
\be
r_{2} \doteq {\mathbb P}\left(X\geq 2\right) \simeq 1-\left[e^{-1/\alpha} \left(1+\frac{1}{\alpha}\right) \right]^{F} . 
\ee
With a reasoning analogous to the one of the previous case, one has that it must be $r_{2}~<~m/M$ for some finite $m$, implying that, for any fixed $M$, $\alpha$ must diverge in the $F~\to~\infty$ limit.

In the nLRSS network of Fig.~\ref{retfig}, $r_{0}$, $r_{1}$, and $r_{2}$ are of similar magnitude, suggesting that all the three small subgraph play a role and that indeed the multiple pairwise featural associations ($r_{2}$) are relevant to the dynamics, including, importantly, the spread of incorrect activity. Using these values, the number of BG modules driven into incorrect stable activity in the first oscillation was estimated well within the correct order of magnitude (the fraction of wrongly active modules is not plotted but was present in the numerical outputs of the simulations).

The formulae for, respectively, $r_{1}$ and $r_{2}$ indicate that, for any arbitrarily large but fixed $M$, $\alpha$ should diverge at least as fast as a power of $F$. It was then tried the function
\be
\alpha \propto F^{\xi}+c , 
\ee
where $c$ and the unstated proportionality factor may be constant or weakly dependent on $F$. Simple asymptotic analysis of Eq.~\ref{r1} indicates that $\xi$ should increase with $F$ for $M$ finite, but arbitrary, and diverging $F$ (indeed, functions in which $\xi$ depended on $M$ but not on $F$ were verified not to produce acceptably good fits simultaneously to the data in Fig.~\ref{PvsM} and in Fig.~\ref{PvsF}). Hence, $\xi$ was also allowed to depend on $F$, looking first for dependence on powers of $\ln(F)$, as suggested by the equations above. In order to conjecture a functional form for the dependence on $M$, it was observed that, because the local dynamics in the network depends on the average number of neighbours per module and because such number stays constant in the limit $M~\to~\infty$, it should be
\be
\lim_{M\to \infty} \alpha = \varphi(F) \in {\mathbb R} . 
\ee
If $\xi$ diverged for $M\to~\infty$ and fixed $F$, the storage capacity would vanish in such limit, which would be at odds with the data. Therefore, the limit should be finite; from analysis and simulations it appears that this number is zero (Fig.~\ref{PvsM}, positive concave function with finite limit) and also that $\varphi(F)$ cannot be a constant (Fig.~\ref{PvsF}). Requiring that $\xi~\to~0$ and $\alpha~\to~\varphi(F)$ for $M~\to~\infty$ leads to 
\be
\alpha \propto \frac{1}{\xi} \left(F^{\xi}-1\right) . 
\ee
Best-fit of the data of Fig.~\ref{PvsM} shows that $\xi$ should depend about exponentially on $M$. 

\section{Table of main symbols and acronyms}
\vspace*{.3cm}\hspace*{-.5cm}
\begin{tabular}[l]{|p{1.4cm}|p{6cm}|} 
\hline 
$M$ & Number of modules in the network \\ 
\hline 
$F$ & Number of features per module  \\
\hline
$P$ & Number of memory patterns stored \\
\hline
$z$ & Mean coordination number of the graph \\
\hline
$\tau$ & Modular sparseness in memory patterns \\
\hline
$\tau t_{1}$ & Correlation of adjacent modules \\
\hline
$(1-\tau)\,t_{0}$ & Anti-correlation of adjacent modules \\
\hline
FG(p) & Foreground of pattern $p$ \\
\hline
BG(p) & Background of pattern $p$  \\
\hline
LRSS & Presence of long-range synaptic scaling \\
\hline
nLRSS & Absence of long-range synaptic scaling \\
\hline
HR & High-robustness of local attractors \\
\hline
LR & Low-robustness of local attractors \\
\hline
$\varrho$ & Fraction of pattern used as a cue \\
\hline
$P_{c}$ & Critical storage-capacity \\
\hline
$a, \beta, \mu, \nu$ & Undetermined best-fit parameters  \\
\hline
\end{tabular}

\end{document}